\documentclass{sig-alternate-10pt}  
\usepackage{listings}
\usepackage{textcomp}
\usepackage{enumitem}
\usepackage{array} 
\usepackage{mathabx, graphicx}
\usepackage{wrapfig}
\usepackage{cite}
\usepackage{xspace}
\usepackage{float} 
\usepackage{pifont}
 
\usepackage{footnote}
  
\usepackage{cleveref}
\usepackage{multirow}
  
\usepackage[margin=1in]{geometry}

\renewcommand{\paragraph}[1]{\vspace{4pt}\noindent\textbf{#1}\xspace}          

\newcommand{\cmark}{\ding{51}}%
\newcommand{\xmark}{\ding{55}}%

\newcolumntype{P}[1]{>{\centering\arraybackslash}p{#1}}

\begin{document}

\pagestyle{empty}

\title{A Closer Look at Variance Implementations \\In Modern Database Systems}

\author{ 
\alignauthor
Niranjan Kamat\hspace{16pt} Arnab Nandi\\
       \affaddr{Computer Science \& Engineering}\\
       \affaddr{The Ohio State University}\\
       \email{\{kamatn,arnab\}@cse.osu.edu}       
}

\maketitle

\begin{abstract}
Variance is a popular and often necessary component of aggregation queries.
It is typically used as a secondary measure to ascertain statistical properties of the result such as its error.
Yet, it is more expensive to compute than primary measures such as \texttt{SUM}, \texttt{MEAN}, and \texttt{COUNT}.

There exist numerous techniques to compute variance. 
While the definition of variance implies two passes over the data, other mathematical formulations lead to a single-pass computation.
Some single-pass formulations, however, can suffer from severe precision loss, especially for large datasets.

In this paper, we study variance implementations in various real-world systems and find that major database systems such as PostgreSQL 9.4 and most likely System X, a major commercial closed-source database, use a representation that is efficient, but suffers from floating point precision loss resulting from catastrophic cancellation. 
We review literature over the past five decades on variance calculation in both the statistics and database communities, and summarize recommendations on implementing variance functions in various settings, such as approximate query processing and large-scale distributed aggregation.
Interestingly, we recommend using the mathematical formula for computing variance if two passes over the data are acceptable due to its precision, parallelizability, and surprisingly computation speed.
\end{abstract}


\section{Introduction}

New large-scale distributed data management and analytics systems are being developed at a rapid pace, with the scalability aspect of computation being their predominant development focus~(excepting \cite{tian2012scalable}). Comparatively lesser efforts have been expended on ensuring numerical correctness and stability of algorithms. While such an approach can result in the queries being answered more quickly, it can also cause the computations to have a higher level of numerical imprecision.

\begin{figure}[ht!]
  \includegraphics[width=\columnwidth]{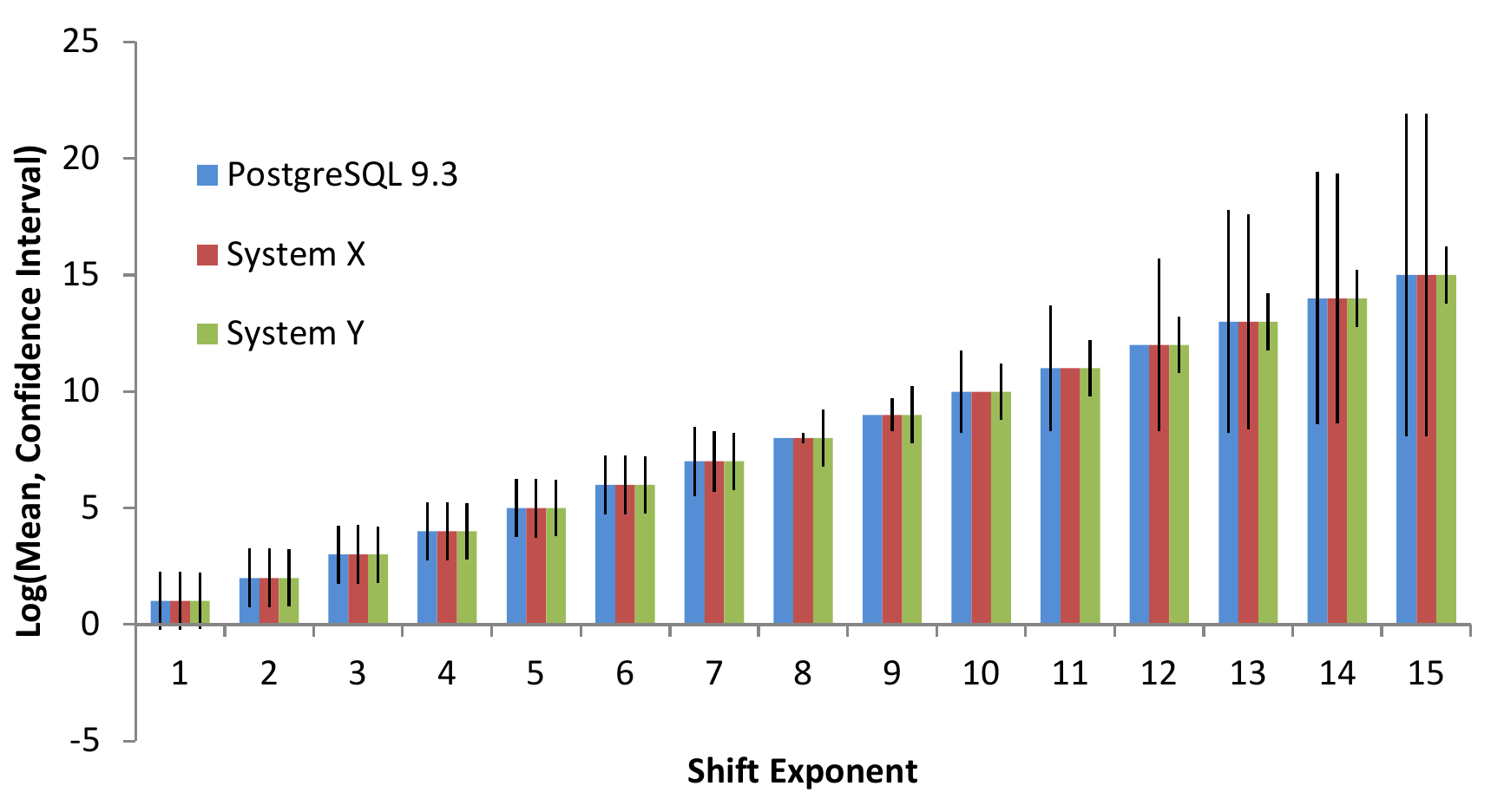}
  \vspace{-25pt}
  \caption{Effect of Variance Error on T-Test Confidence Intervals: As the magnitude of data values increases~(x-axis, true margin of error is kept consistent for each dataset), the mean is expected to increase, and the size of error bars is expected to stay the same. However, PostgreSQL 9.3 and System X error bars ($\alpha = 0.05$) vary widely, while System Y has correct error bars. (100 data points are generated from a \emph{Uniform}$(0,1)$ distribution and shifted using additive shifts of $10^{\emph{\textrm{Shift\ Exponent}}}$ for different values of \emph{Shift Exponent}. A detailed analysis is provided in Section~\ref{intro:visualization}.)}
  \label{fig:error:MOE}
  \vspace{-10pt}
\end{figure}

The concern of achieving numerical stability and precision is pertinent in numerous computational scenarios; it is especially important in variance calculation, which has an ubiquitous presence in large-scale analytics and is known to suffer from precision issues~\cite{higham2002accuracy}.
Variance is an important aggregate function and an essential tool in sampling-based aggregation queries. 
Typically used as a secondary measure, it augments measures such as \texttt{AVERAGE} and provides an insight into the distribution of the data beyond the primary measure.
Computation of variance, however, is susceptible to precision loss when the variance is much smaller than the mean~\cite{chan1983algorithms}.


There exist several techniques to compute variance.
The standard variance formula uses two passes to provide an accurate estimate~(\emph{Two Pass}).
Other techniques using a single pass over data store basic statistics such as count, sum, and sum of squares, due to common perception of \emph{Two Pass} being more expensive due to needing two passes.
One such formula, although fast, is known to suffer from precision loss~(\emph{Textbook One Pass}) due to \emph{catastrophic cancellation}~\cite{higham2002accuracy}, an undesirable effect of a floating point operation that causes the relative error to far exceed the absolute error. Figure~\ref{fig:error:halfCI} demonstrates this problem. As a side note, this problem has been noted to affect calculators as well~\cite{higham2002accuracy}. 

Another formula~(\emph{Updating}), which has been recommended by Knuth~\cite{knuth2014art}, has found a strong foothold in the database community, with numerous implementations citing Knuth in their documentation.
However, this formula is constrained by the fact that it can only incorporate a single data point into the current running estimates. It is unable to combine the estimates from different subsets of data.

Given the rise of large-scale data processing, massive multi-core support and availability of GPUs, it is prudent to consider using representations that can be combined at a larger scale instead of incrementally incorporating a single data point, such as~\emph{Pairwise Updating}. Further,~\emph{Pairwise Updating} is also known to have a better precision, as shown by Chan et al.~\cite{chan1983algorithms} for single precision input~(and, as verified in Section~\ref{experiment}, for double precision as well.) 

\paragraph{Contributions \& Outline:}
\vspace{-5pt}
\begin{itemize}[leftmargin=0cm,itemindent=.5cm,labelwidth=\itemindent,labelsep=0cm,align=left]
  \item We analyze source code for various open source database systems to catalog usage of different variance formulas~(Table~\ref{database:usage}).
  
  \item We experiment with different closed source and open source databases to investigate precision loss issues. We find that precision of PostgreSQL and System X deteriorates the most. After looking at the PostgreSQL source code, we can verify that it uses \emph{Textbook One Pass}, and hypothesize that System X does so as well~(or uses a similar variant).
  
  \item We empirically study the accuracy of the different representations under varying additive shifts and dataset sizes including a hitherto unstudied one, which we call \emph{Total Variance}.

  \item We recommend using \emph{Two Pass} if performing two passes over data is acceptable~(Section~\ref{conclusion}), which seems counter-intuitive, but works due to its computational simplicity.
\end{itemize}

In the next subsection, we look at the adverse effects of imprecise variance calculation. Section~\ref{formulas} presents the different variance representations and their properties. We then detail the representations used by modern databases in Section~\ref{modern-database}. Section~\ref{experiment} lists our analysis of the behavior of the different formulas~(double precision input compared with single precision in Chan et al.~\cite{chan1983algorithms}).
Finally, we conclude with our recommendations for variance representation in current environments.

\subsection{Impact of Variance Calculations}
Due to the pervasive use of variance, a loss of precision can have an impact in a variety of different domains. In the following paragraphs, we look at some use cases where the lack of precision in variance calculation can have adverse consequences.

\paragraph{Incorrect Output:} It is possible to experimentally observe the loss of precision as incorrect output. In order to illustrate the pitfalls in using \emph{Textbook One Pass}, data points were generated from a $Uniform(0,1)$ distribution and shifted by $10^{Shift\ Exponent}$ for \emph{Shift Exponent} varying from $1$ to $14$. 
The variance obtained by using a shift exponent should be expected to be similar to the one without any shift. 
We verify this by adding and subtracting the shift exponent and note that the variance of the resultant dataset was close to the true sample variance.

\begin{figure}[h!]
\vspace{-10pt}
  \includegraphics[width=\columnwidth]{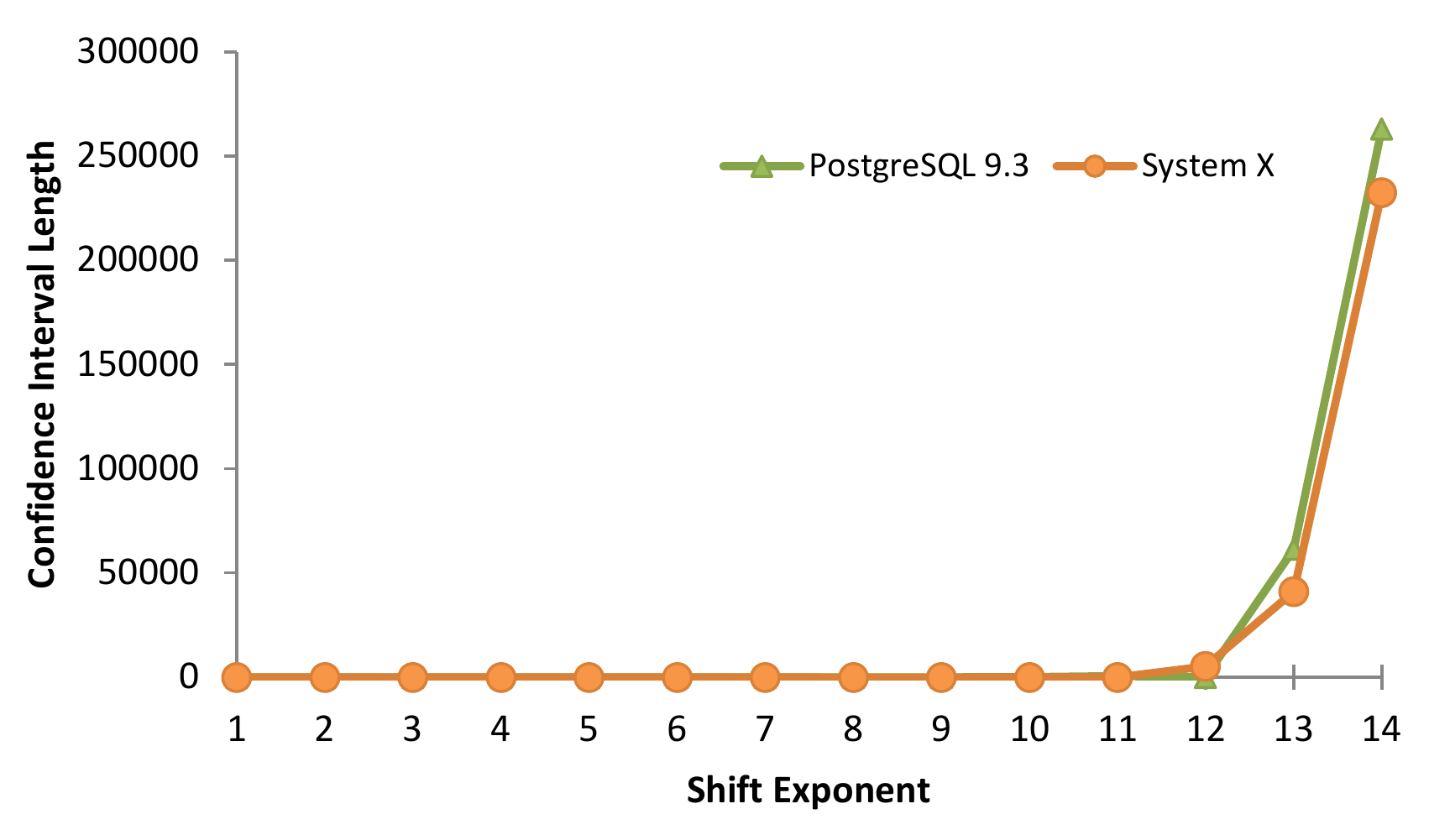}
  \vspace{-25pt}
  \caption{Effect on PostgreSQL and System X: The confidence interval length~($\alpha = 0.05$, \texttt{COUNT} $ = 100$), which is derived from variance, instead of being nearly constant, behaves irrationally due to \emph{Textbook One Pass}. The corresponding PostgreSQL query can be given by \texttt{SELECT $t_{1-\frac{\alpha}{2}} \times$ stddev(column) / sqrt(count(column)) FROM Table}.}
  \label{fig:error:halfCI}
  \vspace{-5pt}
\end{figure}

Figure~\ref{fig:error:halfCI} shows that PostgreSQL 9.3 and System X suffer from variance calculations being susceptible to precision loss since variance should approximately stay the same.
We know that PostgreSQL uses \emph{Textbook One Pass} and the pattern of the erroneous calculations displayed by both of them hints towards System X using it as well.

In contrast, other database systems suffered minor precision loss, as expected~(these results are not shown since they do not add any additional information to the figure). 
It should be noted that System Y was found to be highly immune to precision loss.

\paragraph{Visualization:}
\label{intro:visualization}
Erroneous variance calculation, however small, can have a notable impact on visualizations. 
As a demonstration, we show the results of a repetition of the above experiment in Figure~\ref{fig:error:MOE} and depict the sample mean and the confidence interval.
Due to precision loss, we observe inaccurate results for higher shift values for PostgreSQL 9.3 and System X. While the error bars should be similar, they instead vary widely and inaccurately. 
Error bars for System Y are correctly low throughout.

\paragraph{Negative Variance:}
It is possible for variance to be negative while using \emph{Textbook One Pass} -- a theoretically impossible result~(Table~\ref{expt:one-pass-example}). We observed in the PostgreSQL source code that variance is set to zero, if negative.
Figure~\ref{fig:error:MOE} shows numerous values of 0~(i.e., missing error bars) for PostgreSQL~(shift exponent 8, 9, and 12) and also for System X~(shift exponents 10 and 11), providing evidence of System X employing a similar strategy for handling negative variance values and using \emph{Textbook One Pass}.

\paragraph{Decision support systems:}
As a building block in popular algorithms, flaws in variance implementations can have far-reaching impacts, e.g., in hypothesis testing, which is an integral part of numerous decision support systems. 
Having imprecise or incorrect variance estimates can greatly change the result of hypothesis testing.

\noindent\emph{Loud Failure:} Consider the case of 1 sample 2 tailed t-test with the shift exponent of $8$ using the output of PostgreSQL as given in Figure~\ref{fig:error:MOE}. Let the null hypothesis be as follows:

\emph{$H_0$:} $\mu = 10^{8}+0.483594$~(sample mean)

And the alternate hypothesis as:

\emph{$H_a$:} $\mu \neq 10^{8}+0.483594$ 

The t-statistic can be given by $\frac{\bar{x}-\mu}{\frac{s}{\sqrt{n}}}$, where $\mu$ is the hypothesized mean estimate, $\bar{x}$ is the sample mean, $s$ is the sample standard deviation and $n$ is the sample size. In this case, since $s$ is $0$, the t-test will fail by reporting an error.

\noindent\emph{Silent Failure:} We now look at the more  harmful error of silent failures. Let us consider the sample with shift exponent of $12$ and use the output of System X. Again, let the hypotheses be as follows:

\emph{$H_0$:} $\mu = X$

\emph{$H_a$:} $\mu \neq X$ 

Here, $X$ is the hypothesized population mean. Let $\alpha$~(confidence level) be $0.05$ with the resultant critical value of $1.98$. 
The t-statistic will be $\frac{10^{12} + 0.52-X}{2634.65}$ instead of $\frac{10^{12} + 0.52-X}{0.0278}$. 
If the variance calculation were correct, the range of $X$ for the hypothesis testing to not reject it would have been $[10^{12}+0.475,10^{12}+0.585]$, which is small compared to the now permissible $[10^{12}-5137.03, 10^{12}+5138.08]$.
Thus, we can see that for a large range of $X$, \textbf{the null hypothesis will end up not being rejected without the user any wiser.}


\paragraph{Data Mining:}
Variance is an important tool in statistical analysis and machine learning algorithms such as Gaussian Naive Bayes, or Mixture of Gaussians based algorithms such as background modeling, clustering, or topic modeling.
For example, we found usage of \emph{Textbook One Pass} within a graphics library of the \emph{R} language~\cite{ihaka1996r}. 
Similarly, MADlib~\cite{hellerstein2012madlib} was also found to have a call to the PostgreSQL variance function: thus, an erroneous calculation of variance can extend from the underlying databases to the systems built on top of them.

\begin{table*}
\hfill{}
\begin{tabular}{|c | c | c | c | c | c | c |} 
\hline
Name & Formula & Accuracy & Passes & Storage      &  Parallel\\ 
\hline
\multirow{2}{*}{Two Pass} & $S=\sum_{i=1}^{N}(x_i-\bar{x})^2$ & \multirow{2}{*}{\cmark}    & \multirow{2}{*}{2}      & \multirow{2}{*}{O(1)}         & \multirow{2}{*}{\cmark} \\	&$\bar{x}=\frac{\sum_{x=1}^{N}x_i}{N}$ & & & &\\ 
\hline
Textbook One Pass 	&$S=\sum_{i=1}^{N}x_i^2$ - $\frac{1}{N}(\sum_{i=1}^{N}x_i)^2$ & \xmark   & 1      & O(1)         & \cmark\\ 
\hline

\multirow{2}{*}{Shifted One Pass} & $S=\sum_{i=1}^{N}(x_i-\bar{x})^2 -$ & \multirow{2}{*}{Varies}    & \multirow{2}{*}{1}      & \multirow{2}{*}{O(1)}         & \multirow{2}{*}{\cmark} \\ &$\frac{1}{N}(\sum_{i=1}^N(x_i-\bar{x}))^2$ & & & &\\ 
\hline

\multirow{3}{*}{Pairwise Updating}&$T_{1,m+n} = T_{1,m} + T_{m+1,m+n}$ & \multirow{3}{*}{\cmark}   & \multirow{3}{*}{1}      & \multirow{3}{*}{O(log(N))}         & \multirow{3}{*}{\cmark} \\ 
        &$S_{1,m+n} = S_{1,m} + S_{m+1,m+n} + $              & & & &\\
        &$\frac{m}{n(m+n)}(\frac{n}{m}T_{1,m}-T_{m+1,m+n})^2$ & & & &\\
\hline

\multirow{2}{*}{Updating-YC} & $T_{1,j} = T_{1,j-1} + x_j$ & \multirow{2}{*}{\cmark}   & \multirow{2}{*}{1}      & \multirow{2}{*}{O(1)}         & \multirow{2}{*}{\xmark} \\ 
        &$S_{1,j} = S_{1,j-1} + \frac{1}{j(j-1)}(jx_j - T_{1,j})^2$ & & & &\\
\hline

\multirow{2}{*}{Updating-WWH} & $M_{1,j} = M_{1,j-1} + \frac{x_j - M_{1,j-1}}{j}$ & \multirow{3}{*}{\cmark}   & \multirow{3}{*}{1}      & \multirow{3}{*}{O(1)}         & \multirow{3}{*}{\xmark} \\ 
        \multirow{2}{*}{(Updating)}&$S_{1,j} = S_{1,j-1} + (j-1)\times$ & & & &\\
        &$(x_j - M_{1,j-1})\times(\frac{x_j - M_{1,j-1}}{j})$ & & & &\\
\hline        

\multirow{2}{*}{Total Variance} & $S=\sum_{i = 1}^{groups}n_{i}(m_{i}-\bar{x})^2+$ & \multirow{2}{*}{\cmark}   & \multirow{2}{*}{3}      & \multirow{2}{*}{Varies}         & \multirow{2}{*}{Varies} \\ 
        &$\sum_{i = 1}^{groups}(n_{i}-1)v_{i}$ & & & &\\
\hline
\end{tabular}
\hfill{}
\vspace{-5pt}
\caption{Commonly used Formulas for Variance}
\vspace{-15pt}
\label{intro:common-formulas}
\end{table*}

\section{Different Ways to\\ Calculate Variance}
\label{formulas}

Table~\ref{intro:common-formulas} presents the common variance representations~\cite{chan1983algorithms}. 
We use a similar naming convention to that used by Chan et al.~\cite{chan1983algorithms}.
$S$ stands for the sum of squares.
The sample variance can be given by $\frac{S}{N-1}$, where $N$ is the sample size.
$x_i$ is the $i^{th}$ data point.
$\bar{x}$ is the sample mean.
$M_{m,n}$ is the mean of the data points from indexes $m$ to $n$~(both inclusive).
$T_{m,n}$ is the total of the data points from indexes $m$ to $n$~(both inclusive).
We have also described \emph{Total Variance}, for which we could not find a reference.
In its formula, $n_i$, $m_i$, and $v_i$ represent the count, mean, and variance respectively, of the $i^{th}$ group.

\emph{Textbook One Pass} can be computationally dangerous as the quantities $\sum_{i=1}^{N}x_i^2$ and $\frac{1}{N}(\sum_{i=1}^{N}x_i)^2$ can nearly cancel each other out.
The \emph{Pairwise Updating} formula hierarchically combines pairs of variance values and uses $O(log(N))$ storage while reducing the relative errors from $O(N)$ to $O(log(N))$~\cite{chan1983algorithms}. 
\emph{Updating-YC} represents Youngs and Cramer formula~\cite{youngs1971some} and is essentially identical to \emph{Updating Pairwise} when $m = 1$ or $n = 1$.
The \emph{Updating-WWH} formula refers to the nearly identical formulas used by Welford et al.~\cite{welford1962note}, West et al.\cite{west1979updating}, and Hanson et al. \cite{hanson1975stably} and has similar precision as \emph{Updating-YC}.
We have used the \emph{Updating-WWH} representation for updates using a single data point, and denote it by \emph{Updating}.
Shifting the data by an exact or approximate value of $\bar{x}$~(\emph{Shifted One Pass}) can also result in substantial accuracy gains~\cite{chan1983algorithms}.  

\vspace{4pt}

\subsection{Total Variance}
Since this is the first paper to introduce the \emph{Total Variance} representation, we explain its steps in more details below. In the first pass, which is over the individual tuples, the variance~(using one of the other formulas), mean, and count, of individual groups are computed. The second pass, over the groups thus formed, finds the overall mean of the data. In the third pass, over the groups, the overall variance is then found. 
Since the second and third passes are over the groups obtained as a result of the first pass, and different formulas can be used to compute variance of individual groups, complexity of the overall algorithm can vary widely. 

While this representation is highly parallelizable at the second and third passes, its overall parallelizability is dependent upon the formula used to find variance of individual groups.
Note that this representation is designed for combining variances of different groups and is agnostic to the representation used for individual groups. 
While we have used \emph{Updating} at the group-level in our implementation, it can be replaced by others.

Computing mean of individual groups is a well-researched subject with Tian et al.~\cite{tian2012scalable} providing a good overview. 
We use a single pass algorithm to compute mean of individual groups and to combine means of groups as well. 
To handle a large number of groups, one can look into using an aggregation tree to combine means. 
The usual technique of mean estimation can be used in case the number of groups is large, at the cost of decreased precision.

There does not appear to be a theoretically ideal group size for \emph{Total Variance}, and we could not determine one experimentally either~(Section~\ref{expt:group-size}). 
One natural way of setting group sizes, in distributed execution, is to consider data across different nodes as individual groups. Further, data within a node can be partitioned into equal-sized subgroups, so that each core works on a single subgroup. 


\subsection{Properties of Different Representations}
While Chan et al.~\cite{chan1983algorithms} provide an overview of the accuracy, passes, and storage required for most of the formulas given in Table~\ref{intro:common-formulas}~(other than \emph{Total Variance}), their classification as being distributive, and thus the ability to be parallelized, has not been explicitly listed before, which we do.
In Table~\ref{intro:common-formulas}, the \emph{Storage} column depicts the extra space needed for computing variance, which is above and beyond that needed to store the data itself.



The accuracy of \emph{Shifted One Pass} depends on the accuracy of the estimate of the mean.
\emph{Pairwise Updating} is the only representation giving accurate results while being highly parallelizable and requiring a single pass.
Additionally, as we will see in Section~\ref{experiment}, the precision of \emph{Total Variance} is slightly better than that of \emph{Updating Pairwise}, which typically has the best precision amongst all single pass algorithms. 
As a side note, amongst the different representations, \emph{Two Pass}, \emph{Total Variance} and \emph{Textbook One Pass} are the only ones that can be represented using a standard SQL query.

We note that the error bounds for \emph{Two Pass} are derived by Chan et al. \cite{chan1982updating}, and those for \emph{Textbook One Pass} and \emph{Updating} are provided in \cite{chan1978rounding}. \cite{chan1983algorithms} derives error bounds for \emph{Shifted One Pass}, and conjectures them for \emph{Pairwise Updating}. 
Table 2.1 of \cite{chan1983algorithms} succinctly enumerates them.
Note that Kahan summation~\cite{tian2012scalable, kahan65} can help improve their precision.

\subsection{Data Conditioning}
Data shifting and scaling are immensely useful in improving accuracy of algorithms~\cite{higham2002accuracy}. For example, shifting the data by its mean is the basis for \emph{Shifted One Pass}. Indeed, Chan et al.~\cite{chan1983algorithms} demonstrate the usefulness of shifting by an approximate mean computed using a sample of the data by proving that it reduces the bounds of the condition number. 

Further, numerous techniques such as dividing by the mean or using the log function~\cite{higham2002accuracy} are helpful in improving the accuracy. However, along with requiring additional computational resources these techniques can also worsen the accuracy under malicious datasets~\cite{chan1983algorithms}, and need careful user supervision. 
 
\subsection{Hybrid Formulae}
It is clear that different implementations can be used to find variance of different groups, and combine partial results. 
Indeed, it has been brought to our attention that a commercial system uses the \emph{Updating-YC} formula to compute variance at individual nodes, and combines them using \emph{Pairwise Updating} formula. 
\emph{Total Variance} is a hybrid formula as well, since variance of the groups needs to be computed using one of the other representations. 
This provokes an interesting piece of future work -- choosing different representations at different computation steps, based on factors such as streaming data, numerical precision, data partitioning, time for first result, number of passes permissible. This idea is elaborated upon in Section~\ref{conclusion}.

\subsection{Current Recommendation Guidelines}
Chan et al.~\cite{chan1983algorithms} provide detailed recommendation guidelines on the use of different variance formulas.
They recommend usage of \emph{Pairwise Updating} for combining variances across multiple processors since it reduces the errors and is massively parallelizable if extra $O(log(N))$ space is available. Further, it is also the safest~(least precision loss) algorithm to use within each processor, under the constraint of a single pass.
\emph{Two Pass} provides the best precision amongst all algorithms, but requires two passes.
Based on insights obtained through previous work and our experiments, we provide our guidelines in Section~\ref{conclusion}, which are simple and drastically different from the current guidelines.

\subsection{Extensibility to Other Measures} 
\label{sec:algo:other-measures}
Standard deviation, standard error, and coefficient of variation are important statistical measures, and are based on variance computation. 
As a result, these measures will be affected by the properties of the underlying variance representation.
Similarly, the properties will also extend to any user-defined measure whose variance can be expressed in a closed form as a function of the variance of one of the measure dimensions. For example, for a user-defined measure given by $a$ $*$ \texttt{AVG}$(Agg) + b$, where $a$ and $b$ are constants and $Agg$ is a measure dimension, the variance of the measure can be given in closed form as $a^2*$\texttt{VARIANCE}$(Agg)$. 
Note that obtaining a closed form solution to the variance of holistic or complex measures is not always possible, with bootstrapping being a popular choice for variance estimation~\cite{kleiner2013general}.

\section{Variance Implementations \\in Modern Database Systems}
\label{modern-database}

Given the variety of variance formulas, we now survey various open source databases to find out which formulas are used by them to compute variance. Based on our experiments, we also conjecture about two closed source databases. Table~\ref{database:usage} lists the formula used in each database system.

 \vspace{-5pt}
\begin{table}[h!]
\begin{tabular}{|P{55pt} |c|} 
\hline
Database                    & Formula \\ 
\hline
PostgreSQL 9.4.4 	        & \multirow{2}{*}{Textbook One Pass}\\ 
\hline
MySQL 5.7   	            & Updating\\ 
\hline 
Impala 2.1.5                & Updating Pairwise\\ 
\hline
Hive 1.2.1                  & Updating Pairwise\\ 
\hline
Spark 1.4.1     	        & Updating Pairwise\\
\hline
SQLite 	                    & No Variance Support\\  
\hline
System X                    & Textbook One-pass\emph{~(Conjecture)}\\ 
\hline
\multirow{2}{*}{System Y}   &Higher precision variables\emph{~(Guess).}\\ 	
				            &\emph{Cannot conjecture about formula}\\ 
\hline
\end{tabular}
\vspace{-10pt}
\caption{Variance Implementations in Modern Databases}
\vspace{-5pt}
\label{database:usage}
\end{table}


PostgreSQL uses \emph{Textbook One Pass} and is thus susceptible to precision loss.
MySQL uses Knuth's modification~\cite{knuth2014art} of Welford's updating formula. 
Therefore, it can only process a single additional data point, and cannot avail of the possible parallelization.
Spark 1.4.1 and Impala 2.1.5, on the other hand, use a modified version of \emph{Updating Pairwise}. 

Although the source code for System X is not available, we conjecture that it uses \emph{Textbook One Pass} as its precision behavior was similar to that of PostgreSQL. 
System Y was found to have the best precision. We hypothesize that it uses higher precision variables, but cannot make any conjecture about the exact representation.


\section{Experimental Analysis}
\label{experiment}
Chan et al.~\cite{chan1983algorithms} have looked at the precision of different algorithms using single precision input. 
We present the precision results using double precision input. 
We also evaluate the precision of \emph{Total Variance}.
In addition, we look at the precision in the variance calculation offered by the different databases.
We also present the execution times of different algorithms on data sizes up to $100$ million tuples.
The presented results are the average over $100$ runs. 
Results from Section~\ref{sec:algo:applicability} till Section~\ref{expt:in-action} were performed using Ubuntu 14.04.05 LTS with a 4 core, 2.4 GHz Intel CPU, with 16 GB RAM, and 256 GB SSD storage, on a single execution thread.
To look at the parallelization speedups, which are possible for some representations, Section~\ref{expt:multi-threading} provides multi-threading-based results. 


\vspace{5pt}
\noindent \textbf{Dataset:}
Although numerous benchmarks exist to evaluate the accuracy of numerical algorithms, they are constrained by the fact that their the dataset sizes are quite limited. For example, the biggest dataset in the NIST StRD~\cite{rogers1998strd} benchmark consists of $5000$ points. Furthermore, for this dataset, the mean is not significantly  larger than the standard deviation ($\mu = 4.5348$, $\sigma = 2.8673$). Therefore, in a similar vein as Tian et al.~\cite{tian2012scalable}, who generated simulated datasets inspired by NIST StRD, we created synthetic datasets of different sizes with double precision from $Uniform(0,1)$, with the resulting variance of $\frac{1}{12}$. 
These samples were shifted by adding values ranging from $10^1$ to $10^{15}$.

\subsection{Impact of Shift}
\label{sec:algo:applicability}

\vspace{-15pt}
\begin{figure}[h!]
\includegraphics[width=\columnwidth]{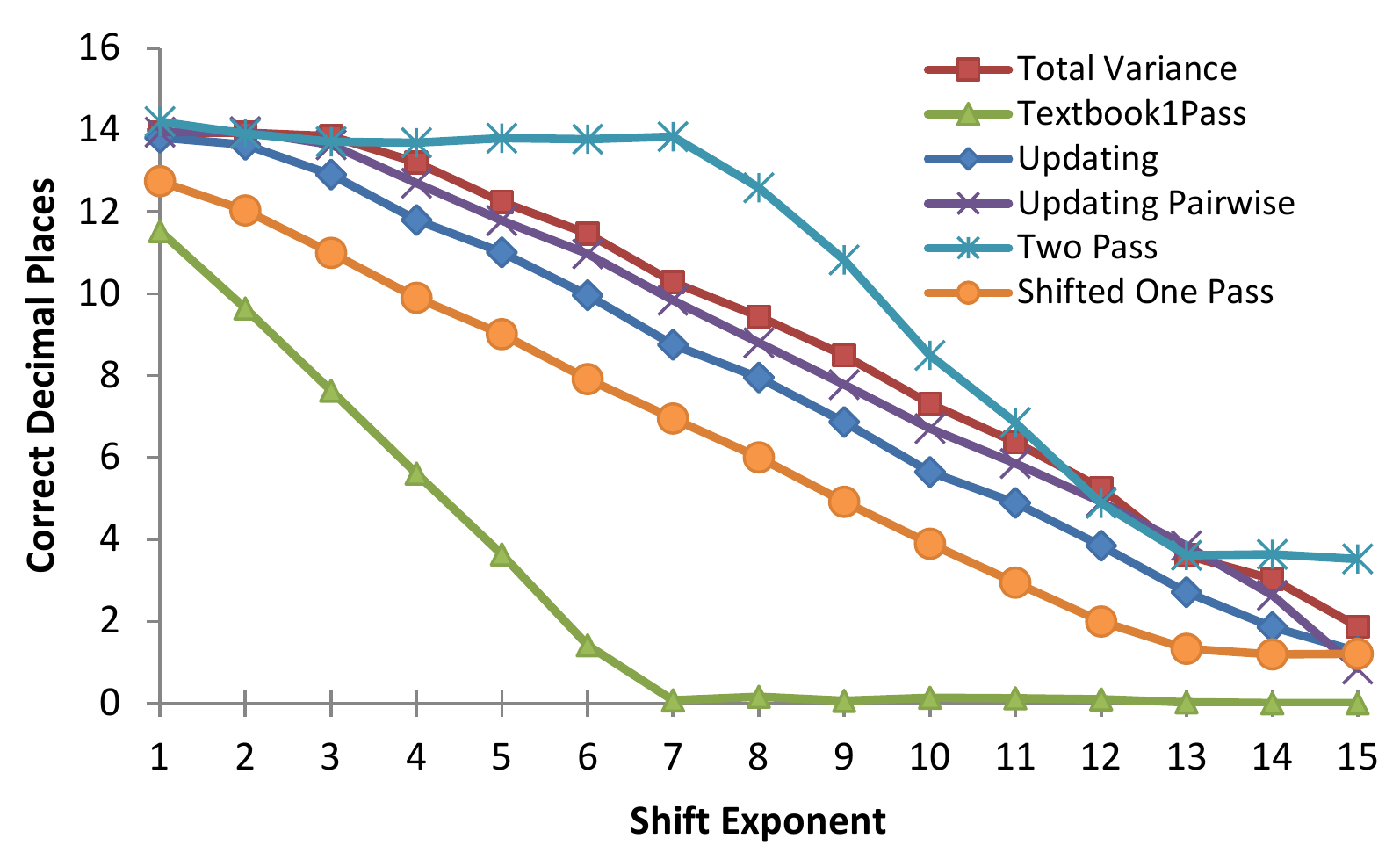}
\vspace{-25pt}
\caption{Impact of Increasing Shift on Precision: With increasing shift exponent, all representations experience precision loss -- though some more severely than others.}
\vspace{-10pt}
\label{fig:error:precision}
\end{figure}
Numerical precision was evaluated using varying additive shift exponents, over a dataset of size $10000$. 
Group size was set to $10$ for \emph{Total Variance}.
We present our findings in Figure~\ref{fig:error:precision}, where Y-axis represents the number of correct decimal digits~(non-fractional part of the result was $0$). 
We found the results to be as expected~\cite{chan1983algorithms}, with \emph{Two Pass} having the best precision, and \emph{Textbook One Pass} being clearly impacted by the increasing shift exponent.

\subsection{Impact of Data Size}
\label{sec:expt:data-size}

\begin{figure}[h!]
\includegraphics[width=\columnwidth]{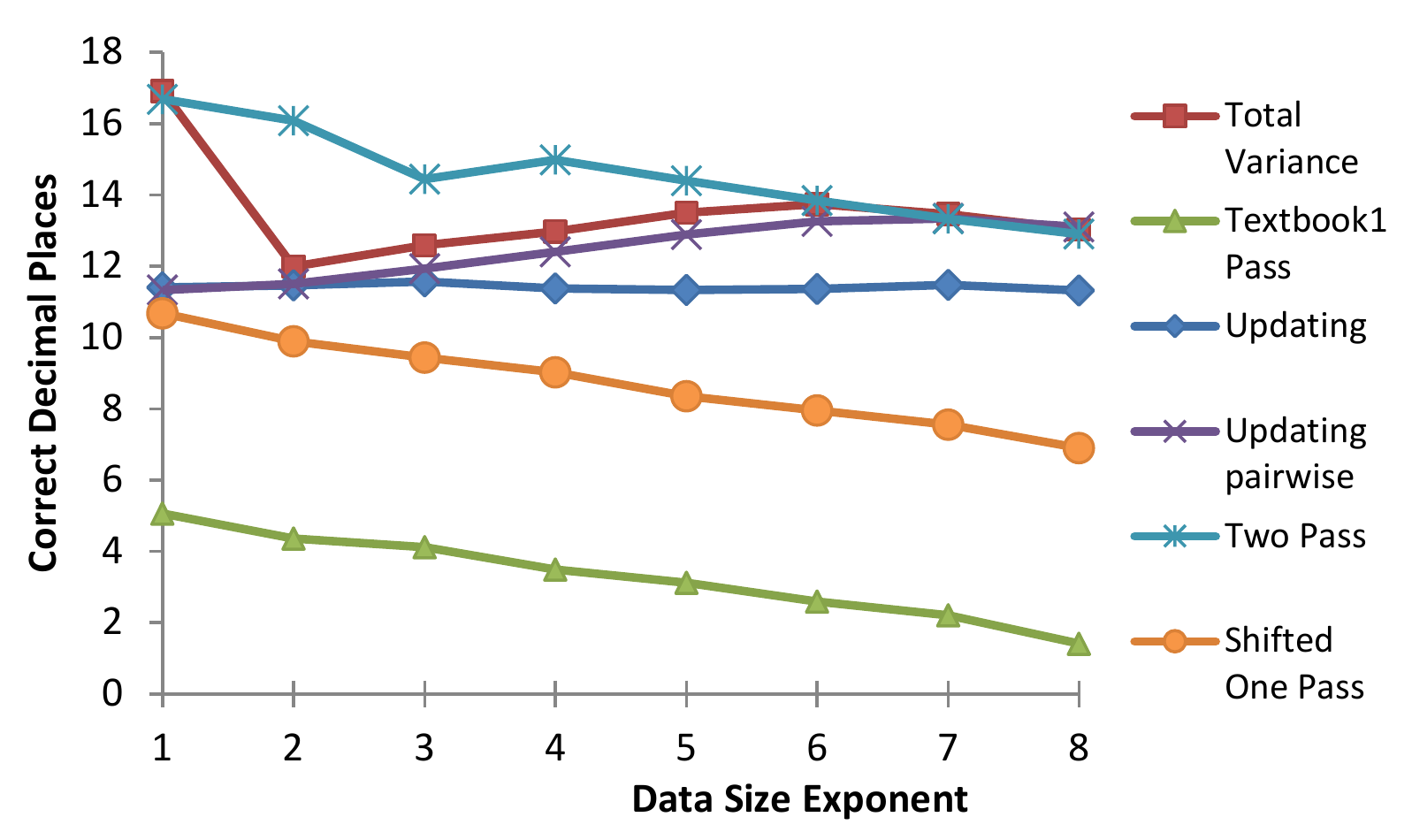}
\vspace{-25pt}
\caption{Impact of Increasing Data Size on Precision: Precision generally decreases with increasing dataset size, with exceptions of \emph{Total Variance} and \emph{Updating Pairwise}.}
\vspace{-15pt}
\label{fig:error:data-size}
\end{figure}

Since precision errors typically accumulate, we used datasets of sizes ranging from $10$ to $100$ million. The shift was set at $10^5$.
We can see from Figure~\ref{fig:error:data-size} that as expected, in most cases the precision worsens with increasing data size. \emph{Two Pass} again outperforms other algorithms. \emph{Textbook One Pass} shows consistently worst precision.

Counter-intuitively, the precision of \emph{Total Variance} and \emph{Updating Pairwise} was found to increase with the data size exponent from 2 to 6. We are unable to conjecture the reason behind this behavior. The precision error for \emph{Updating Pairwise} increases as $O(log(n))$, while that for others~(except \emph{Total Variance}) increases as at least $O(n)$~\cite{chan1983algorithms}, where $n$ is the data size. 
Therefore, while we can expect the error in \emph{Updating Pairwise} to not increase at the same rate as other algorithms, the error decrease is unexpected. 
In the absence of theoretical error bounds for \emph{Total Variance}, we cannot hypothesize about the possible causes for its behavior.
To ensure there were no irregularities, the experiment was repeated multiple times with similar results.

\subsection{Impact of Shift on Different Databases}

\vspace{-15pt}
\begin{figure}[h!]
\includegraphics[width=\columnwidth]{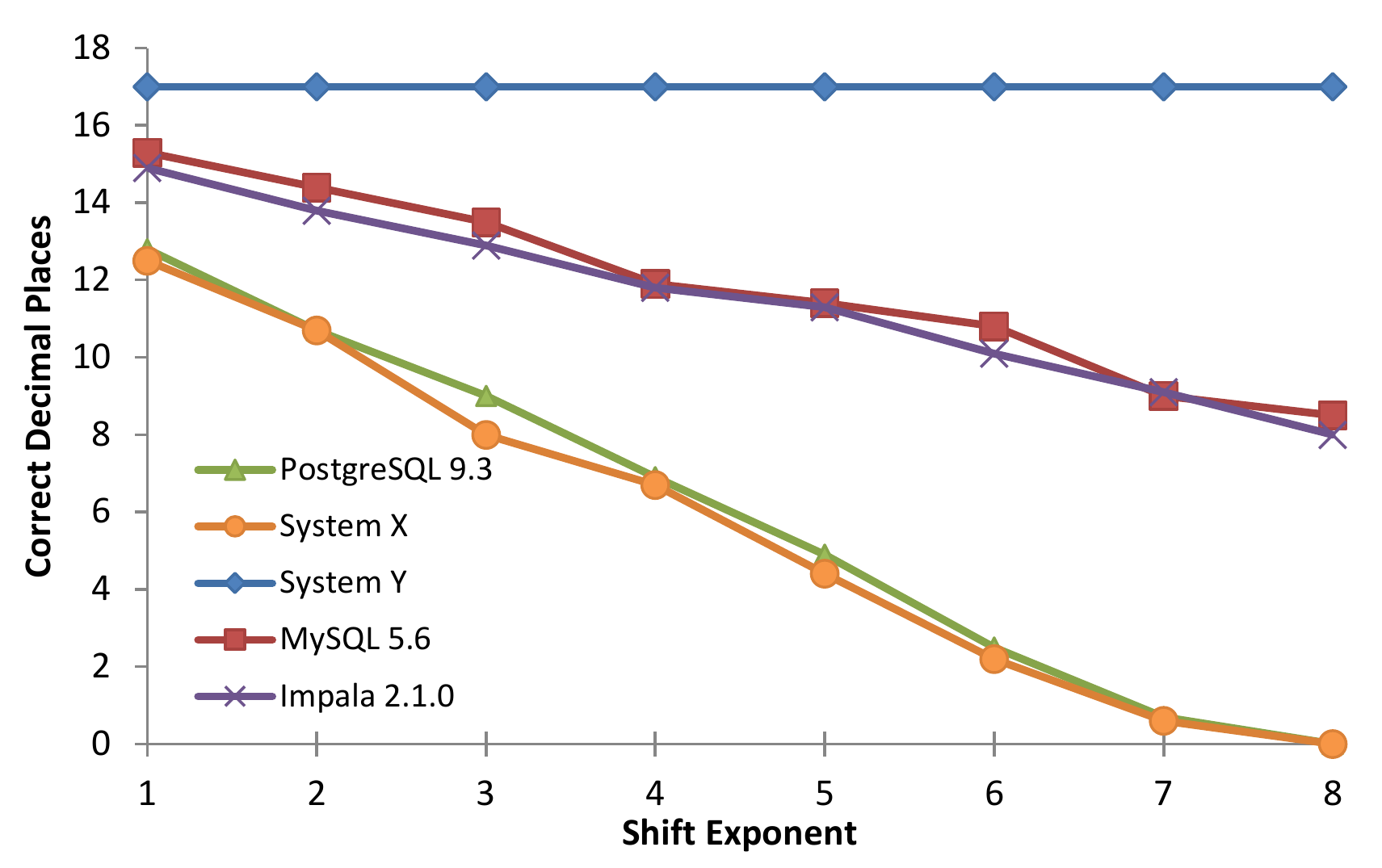}    
\vspace{-25pt}
\caption{Impact of Shift on Databases: Databases follow precision patterns that are expected from their variance formulas.}
\vspace{-15pt}
\label{fig:error:correct-decimal}
\end{figure}
We  look at variance precision for different databases under varying additive shifts, for similar datasets, which are prone to precision errors. 
We took efforts to ensure different systems have similar data types.
100 points were chosen from a $Uniform(0, 1)$ distribution.
Figure~\ref{fig:error:correct-decimal} shows that precision loss follows a similar pattern in  System X and PostgreSQL. Impala and MySQL have a similar error profile as well.

\begin{table*}
\hfill{}
\vspace{30pt}
\begin{tabular}{|c | c | c | c | c |} 
\hline
Shift    & $mantissa(S_1)$                 & $mantissa(S_2)$            & $S=S_1-S_2$         & variance\\ 
\hline
None     & 0xa7677ed386b82                 & 0x3f74ce8319d49            & 831.5840227247941 & 0.08316671894437384\\
\hline
1        & 0xda6e1ccdb823                  & 0xd72e874b34ca             & 831.5840227215085 & 0.08316671894404526\\ 
\hline
2        & 0x8156ee01176cb                 & 0x81561e1bb6cb4            & 831.5840228646994 & 0.08316671895836578\\ 
\hline
3        & 0x2a531c0d87ff3                 & 0x2a531a6dbd3c9            & 831.5840644836426 & 0.08316672312067633\\ 
\hline
4        & 0xd1b557c3f3080                 & 0xd1b557bd73dc9            & 831.5848388671875 & 0.08316680056677543\\ 
\hline
5        & 0x6bcd32f7f2a8c                 & 0x6bcd32f7e5a78            & 832.3125          & 0.08323957395739574\\ 
\hline
6        & 0x1c37a6532f3c2                 & 0x1c37a6532f25c            & 716.0             & 0.07160716071607161\\ 
\hline
7        & 0xbc16d9663a96c                 & 0xbc16d9663a8ef            & 16000.0           & 1.6001600160016\\ 
\hline
8        & 0x5af1d7c632dda                 & 0x5af1d7c632df7            & -475136.0         & -47.518351835183516\\ 
\hline
\end{tabular}
\vspace{-36pt}
\hfill{}
\caption{Example of Application of \emph{Textbook One Pass}}
\vspace{-10pt}
\label{expt:one-pass-example}
\end{table*}

\subsection{Single-Threaded Execution Speed}
\label{expt:speed:local:single-thread}

\begin{figure}[h!]
\includegraphics[width=\columnwidth]{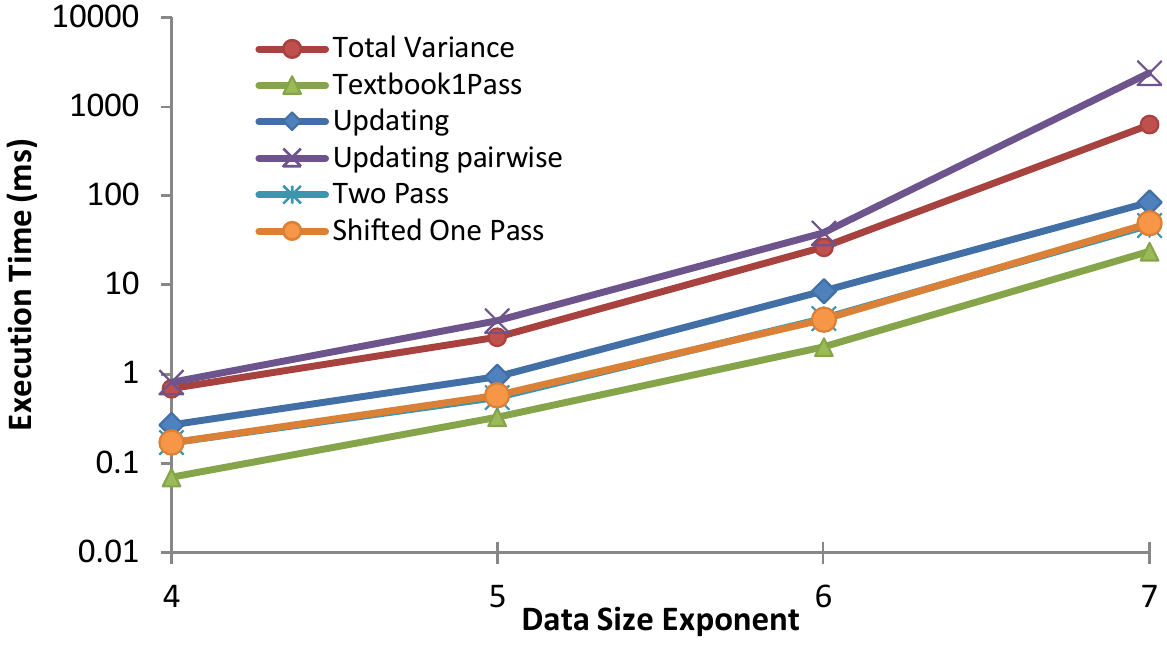}
\vspace{-15pt}
\caption{Single-Threaded Execution Speed: Though \emph{Two Pass} requires 2 passes over data, it provides results faster than other algorithms, with exception of \emph{Textbook 1 Pass}, which has the least numerical precision.}
\vspace{-10pt}
\label{fig:speed:single-thread}
\end{figure}
We also looked at the execution time of different algorithms with increasing data size.
Results with lower data sizes have not been presented due to the computation taking minimal time.
This experiment presented us with interesting results.
Surprisingly, there was no discernible difference in execution time between \emph{Two Pass} and \emph{Shifted One Pass}.
\emph{Textbook One Pass} was the only algorithm that took lesser time than \emph{Two Pass}.
We attribute the low execution time of \emph{Two Pass} to simplicity of its computation.
Due to superior accuracy, least execution time after error-prone \emph{Textbook One Pass}, and ease of implementation and parallelization, we suggest that \emph{Two Pass} should be the algorithm of choice if performing two passes over the data is acceptable. 

\subsection{\mbox{Impact of Group Size on Precision}}
\label{expt:group-size}
\begin{figure}[h!]
\includegraphics[width=\columnwidth]{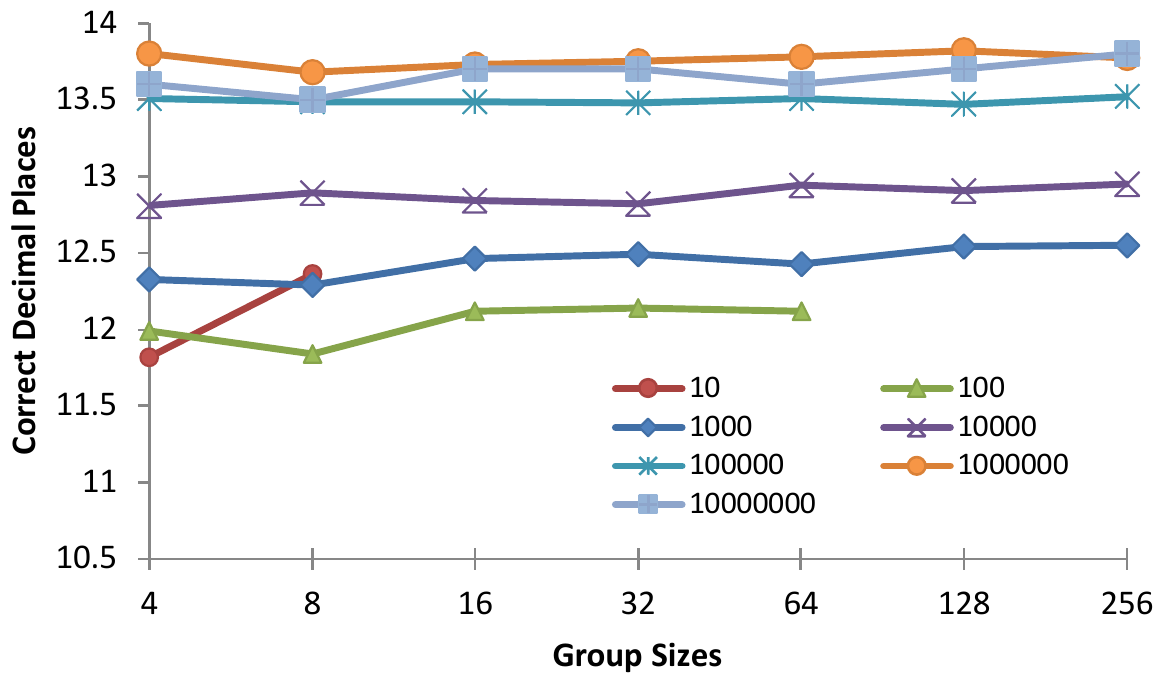}
\vspace{-20pt}
\caption{Impact of Group Size on Precision: Increasing group size improves precision slightly for some data sizes, although there does not exist a clear relationship between precision and group size.}
\vspace{-15pt}
\label{fig:group-sizes}
\end{figure}
Since group size is an integral component of our \emph{Total Variance} algorithm, we looked at the effect different group sizes have on precision.
Figure~\ref{fig:group-sizes} shows that there does not exist any clear relationship between them, though precision increased in a majority of cases with increasing group size.
Thus, there does not appear to be any ideal group size from the perspective of precision.
We also note that there did not exist any significant differences in the execution time for varying group sizes.

\subsection{Textbook One Pass in Action}
\label{expt:in-action}
To further illustrate catastrophic cancellation occurring in \emph{Textbook One Pass}, Table~\ref{expt:one-pass-example} presents the corresponding mantissa of the two expressions that compose it. We consider a random sample of size $10000$ generated from $Uniform(0,1)$, and shift it by exponents ranging from $1$ to $7$.
Note that \emph{Textbook One Pass} calculates the sum of squares as $S=S_1 -S_2$, where 
$S_1=\sum_{i=1}^{N}x_i^2$ and $S_2 =\frac{1}{N}(\sum_{i=1}^{N}x_i)^2$.
We can see that an increasing number of bits in the mantissa of $S_1$ and $S_2$ become equal, until all precision is lost for the shift exponent of 6.

\subsection{Multi-Threaded Execution Speed}
\label{expt:multi-threading}

\begin{figure}[h!]
\includegraphics[width=\columnwidth]{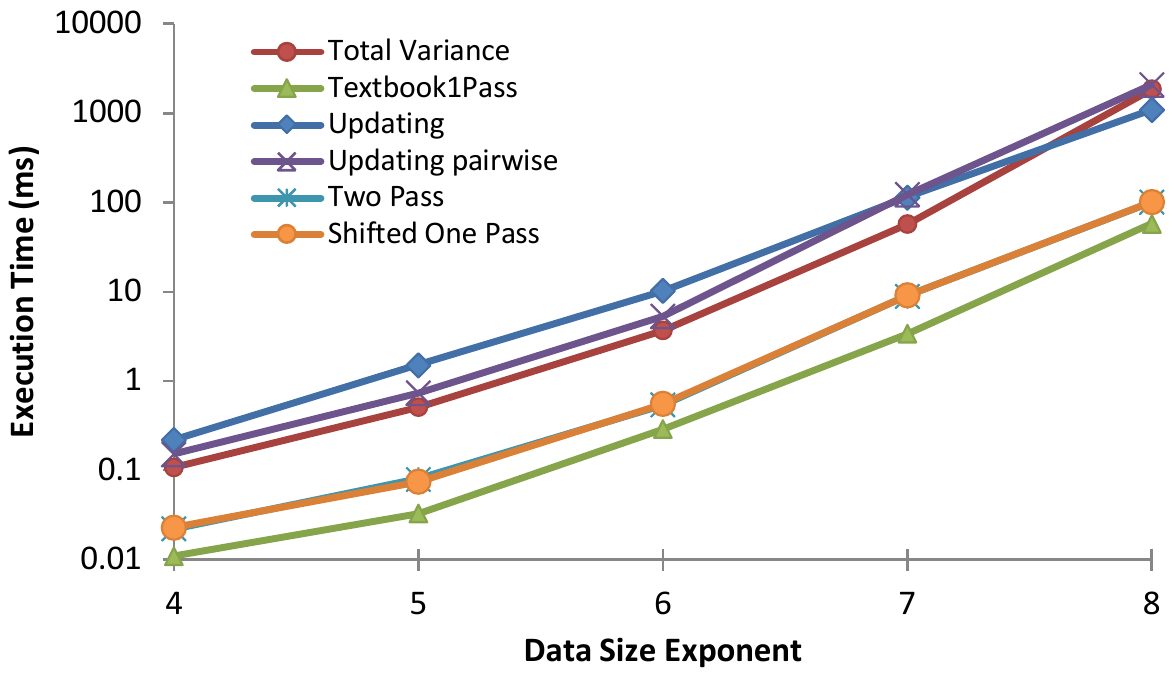}
\vspace{-20pt}
\caption{Multi-Threaded Execution Speed: \emph{Updating} cannot avail of parallelism, while other algorithms can. \emph{Two Pass} again provides results quicker than \emph{Updating}, \emph{Updating Pairwise}, and \emph{Total Variance}.}
\vspace{-15pt}
\label{fig:speed:multi-thread}
\end{figure}
To determine the possible speedups due to parallel execution, the algorithms were parallelized and run on an Ubuntu Linux 14.04.1 LTS system with a 48 core 2.4 GHz Intel Xeon CPU, with 256 GB memory, and a 500 GB disk.
With the exception of \emph{Updating}, other representations were able to benefit from parallelism.
We can again observe that \emph{Two Pass} has similar execution time as \emph{Shifted One Pass}, with only \emph{Textbook One Pass} taking lesser time.
Thus, in both single-threaded and multi-threaded environments, \emph{Two Pass} performed exceedingly well.

We note that there were only minor changes in precision due to small modifications being added to them for parallelization. 
Further, in a similar fashion as Section~\ref{expt:group-size}, varying group sizes in the \emph{Total Variance} representation did not result in significant difference in precision or execution time.

\section{Conclusion \& Recommendations}
\label{conclusion}
Floating point precision can cause information loss in both data measurement as well as data storage. This problem is further exacerbated to varying degrees by different variance calculation formulae.

Precision issues associated with \emph{Textbook One Pass} have been well documented. However, we have seen that databases such as PostgeSQL and likely System X still use it. 
We recommend from the perspective of safety to discontinue its usage. Though there might be arguments for its continued usage after warning the users in certain scenarios, the arguments against it far outweigh the speedup benefit and its ease of implementation. Although error inherently exists in approximate query processing, numerical precision errors are easy to eliminate and hard to apportion and therefore should be avoided whenever possible. 
Hence, we recommend to the designers of databases, and statistics and analytics packages, to discontinue its usage. Further, it would be wise for users to perform a sanity check using experiments similar to those given in Section~\ref{sec:algo:applicability}.



Previous work has recommended \emph{Pairwise Updating} from the perspective of precision, speed, and parallelizability~\cite{chan1983algorithms}. 
However, we have seen from our experiments of up to 100 million data points, that the most accurate algorithm, \emph{Two Pass}, takes lesser time than \emph{Updating}, \emph{Updating Pairwise}, and \emph{Total Variance}.
Further, it takes around the same amount of time as \emph{Shifted One Pass}, which relies on mean estimation. 
\emph{Two Pass} is also easy to implement and parallelize. Therefore, in the case that  \textbf{performing two passes over the data is acceptable, \emph{Two Pass} should be the preferred algorithm}.
Determining whether two passes are acceptable, however, is a nuanced decision. 
When the data fits in memory, performing two passes over the data is clearly acceptable as all representations will incur the identical data read I/O cost.
When the data cannot fit in memory, summing up the estimated I/O and computation times can help determine whether \emph{Two Pass} will need the least amount of time, in which case it should be chosen.


\textbf{In other cases, i.e., whenever \emph{Two Pass} is not estimated to require the least execution time,
there does not exist a clear winner}, due to different algorithms having different strengths and weaknesses. 
\emph{Updating} provides faster results at lower precision, compared with \emph{Updating Pairwise}, without needing additional memory. 
\emph{Updating Pairwise} is parallelizable, whereas \emph{Updating} is not.
While \emph{Shifted One Pass} provides quick results, its accuracy is dependent on correctness of the mean estimate.
\emph{Total Variance} has good accuracy, although it takes longer to execute, and is dependent on the algorithm used to compute group statistics, while also needing multiple passes.
Hence, there does not exist any algorithm that dominates every other algorithm, resulting in there not being a clear choice.
We can thus see that a query planner that devises hybrid formulas, while taking the data distribution, estimated I/O and computation costs, and the overall strengths and weaknesses of different algorithms into consideration, appears to be an important and ideal piece of future work.

\section{Acknowledgment}
We acknowledge the generous support of U.S. National Science Foundation under awards IIS-1422977 and CAREER IIS-1453582.
We would also like to thank the SIGMOD RECORD reviewer for their insightful and helpful comments, which greatly improved our paper.

\bibliographystyle{abbrv}
\bibliography{variance}

\begin{thebibliography}{10}

\bibitem{chan1982updating}
T.~F. Chan et~al.
\newblock {Updating Formulae and a Pairwise Algorithm for Computing Sample
  Variances}.
\newblock {\em COMPSTAT}, 1982.

\bibitem{chan1983algorithms}
T.~F. Chan et~al.
\newblock {Algorithms for Computing the Sample Variance: Analysis and
  Recommendations}.
\newblock {\em Am. Stat.}, 1983.

\bibitem{chan1978rounding}
T.~F. Chan and J.~Lewis.
\newblock {Rounding Error Analysis of Algorithms for Computing Means and
  Standard Deviations}.
\newblock {\em JHU, TR}, 1978.

\bibitem{hanson1975stably}
R.~J. Hanson.
\newblock {Stably Updating Mean and Standard Deviation of Data}.
\newblock {\em ACM}, 1975.

\bibitem{hellerstein2012madlib}
J.~M. Hellerstein et~al.
\newblock {The MADlib Analytics Library: or MAD Skills, the SQL}.
\newblock {\em VLDB}, 2012.

\bibitem{higham2002accuracy}
N.~J. Higham.
\newblock {Accuracy and Stability of Numerical Algorithms}.
\newblock {\em SIAM}, 2002.

\bibitem{ihaka1996r}
R.~Ihaka et~al.
\newblock {R: A Language for Data Analysis and Graphics}.
\newblock {\em {J. Comp. Graph. Stat.}}, 1996.

\bibitem{kahan65}
W.~Kahan.
\newblock {Further Remarks on Reducing Truncation Errors}.
\newblock {\em ACM}, 8(1):40, 1965.

\bibitem{kleiner2013general}
A.~Kleiner et~al.
\newblock {A General Bootstrap Performance Diagnostic}.
\newblock {\em SIGKDD}, 2013.

\bibitem{knuth2014art}
D.~E. Knuth.
\newblock {\em Art of Computer Programming, Volume 2: Seminumerical
  Algorithms}.
\newblock 2014.

\bibitem{rogers1998strd}
J.~Rogers et~al.
\newblock {StRD: Statistical Reference Datasets for Testing the Numerical
  Accuracy of Statistical Software}, 1998.

\bibitem{tian2012scalable}
Y.~Tian et~al.
\newblock {Scalable and Numerically Stable Descriptive Statistics in SystemML}.
\newblock {\em ICDE}, 2012.

\bibitem{welford1962note}
B.~Welford.
\newblock {Note on a Method for Calculating Corrected Sums of Squares and
  Products}.
\newblock {\em Technometrics}, 1962.

\bibitem{west1979updating}
D.~West.
\newblock {Updating Mean and Variance Estimates: An Improved Method}.
\newblock 1979.

\bibitem{youngs1971some}
E.~A. Youngs et~al.
\newblock {Some Results Relevant to Choice of Sum and Sum-of-product
  Algorithms}.
\newblock {\em Technometrics}, 1971.

\end{thebibliography}
 
\section{APPENDIX}
\subsection{Total Variance Derivation}
Suppose the dataset $D$ contains N data points, with the $i^{th}$ data point having the value $x_i$. Sample variance of the entire dataset can be given by $v = \frac{1}{N - 1} \sum_{i = 1} ^ N (x_i - \bar{x})^2$, and the sample mean by $\bar{x} = \frac{1}{N} \sum_{i = 1} ^ N x_i$.
Let $D$ consist of $K$ separate groups, with the $i^{th}$ group $D_i$ consisting of $n_i$ data points. The mean,~$m_i$, and variance,~$v_i$, of the $i^{th}$ group can then be given respectively by 
\begin{flalign}
	&m_i = \frac{1}{n_i} \sum_{j \in D_i}  x_j & \nonumber\\
	& v_i = \frac{1}{n_i - 1} \sum_{j \in D_i} (x_j - m_i)^2 & \nonumber
\end{flalign}

\noindent The sample variance of $D$ can then be broken up as
\begin{flalign}
	&v = \frac{1}{N - 1} \sum_{i = 1} ^ N (x_i - \bar{x})^2& \nonumber
\end{flalign}
\begin{flalign}
	& = \frac{1}{N - 1} \sum_{i = 1} ^ K \sum_{j \in D_i}   (x_j - \bar{x})^2& \nonumber
	\\& = \frac{1}{N - 1} \sum_{i = 1} ^ K \sum_{j \in D_i}   (x_j - m_i + m_i - \bar{x})^2& \nonumber
	& = \frac{1}{N - 1} \sum_{i = 1} ^ K \sum_{j \in D_i}   [ (x_j - m_i)^2 + 2(x_j - m_i)(m_i - \bar{x}) + (m_i - \bar{x})^2]& \nonumber
	\\& = \frac{1}{N - 1} \sum_{i = 1} ^ K [ \sum_{j \in D_i}(x_j - m_i)^2 + 2(m_i - \bar{x})\sum_{j \in D_i}(x_j - m_i) & \nonumber\\
	& \qquad \qquad + \sum_{j \in D_i}(m_i - \bar{x})^2] & \nonumber
	\\& = \frac{1}{N - 1} \sum_{i = 1} ^ K [ (n_i - 1) s_i^2 + 2(m_i - \bar{x})(n_im_i - n_im_i) + n_i(m_i - \bar{x})^2]& \nonumber
	\\& = \frac{1}{N - 1} \sum_{i = 1} ^ K [ (n_i - 1) v_i + n_i(m_i - \bar{x})^2]& \nonumber
\end{flalign}

\end{document}